\documentclass[a4paper,11pt]{article}
\usepackage{jinstpub} 
\usepackage{lineno}

\title{\boldmath
PhotonPix: Single-Photon Detector with 10 ps timing precision and high dynamic range
}

\author[a,1]{D.A. Orlov,\note{Corresponding author.} 
\emailAdd{d.orlov@exosens.com}
}
\author[b]{Y. Prokazov,}
\author[b]{E. Turbin,} 
\author[a]{and E. Kernen}
\affiliation[a]{Exosens,\\Dwazziewegen 2, 9301ZR Roden, The Netherlands}
\affiliation[b]{Photonscore GmbH,\\Klosterwuhne 42, 39124 Magdeburg, Germany}

\abstract{
A plug-and-play PhotonPix\texttrademark\ single-photon detector with a logical signal output is developed 
for applications requiring ultimate timing precision down to 10 ps over a wide dynamic photon flux range. 
The heart of the detector is an Exosens Fast Timing Microchannel Plate Photomultiplier (FT MCP-PMT) 
with a large 8 mm diameter sensitive area, which can accommodate various Hi-QE photocathodes optimized 
for high quantum efficiency (QE) and low dark rates. 
The detector dead time, timing accuracy, and counting efficiency of the PhotonPix\texttrademark\ 
are measured and analyzed over a wide dynamic photon flux range up to about 1 GHz in burst mode 
and up to 100~MHz in continuous operation mode.
}


\begin{document}
\maketitle
\flushbottom

\section{Introduction}
\label{sec:intro}
Time-correlated single-photon counting (TCSPC) is widely used in fields such as light detection and ranging (LiDAR), medical imaging, high-energy particle detection, astrophysics, materials science, quantum key distribution, fiber-optic communication, and quantum information processing. In addition to high temporal resolution, typically a few tens of picoseconds, a wide dynamic range of detectable photon flux is often required. In most practical cases, the maximum average rate does not exceed a few MHz; however, burst-mode counting may demand rates up to a GHz. For instance, in calorimetry, Cherenkov detection, or scintillator readout, the system must register single photons from low-energy events while tolerating occasional bursts from particle showers or scintillation cascades. Consequently, the detector operates predominantly at low photon flux, interrupted by sporadic high-intensity bursts, therefore a short dead time is essential to mitigate pile-up and saturation. Similar requirements arise in LiDAR, TCSPC-based fluorescence lifetime imaging (FLIM), optical communication, and astrophysics; and some applications push the limits of high-throughput single-photon detection even further, for example, in stellar intensity interferometry measurements \cite{Spie Erlangen}.

While the FT MCP-PMT can be considered as an ideal detector for photon counting time resolved applications in the photon flux range up to several MHz, at higher rates it can face different challenges. For example, output signal waveform (e.g., pulse width and ringing) can limit the achievable dead time, reducing the maximum detectable rate. Also, MCP saturation can further decrease signal amplitude, effecting the counting efficiency and timing performance. In addition, high average-rate operation can also shorten photocathode lifetime. On the electronics side, the pulse processing, bandwidth, and counting capacity, pose significant challenges at high rates. Thus, achieving ultimate timing resolution while maintaining a wide photon-flux dynamic range is challenging for both the detector and front-end electronics. 

In this paper, we present the PhotonPix\texttrademark\ detector, an optimized integration of a fast-timing FT8 MCP-PMT sensor with its advanced front-end electronics. This plug-and-play solution provides a NIM-logic signal output with timing resolution down to 10~ps and a wide dynamic range. At low rates, detection is limited by the sensor dark events, typically below 10~cps, while at high fluxes the maximum detectable rate is constrained by a dead time of approximately 1.6~ns. 

\section{Fast-Timing FT-8 MCP-PMT}
\label{sec:FT8}

\begin{figure}[htbp]
\centering
\includegraphics[width=.47\textwidth]{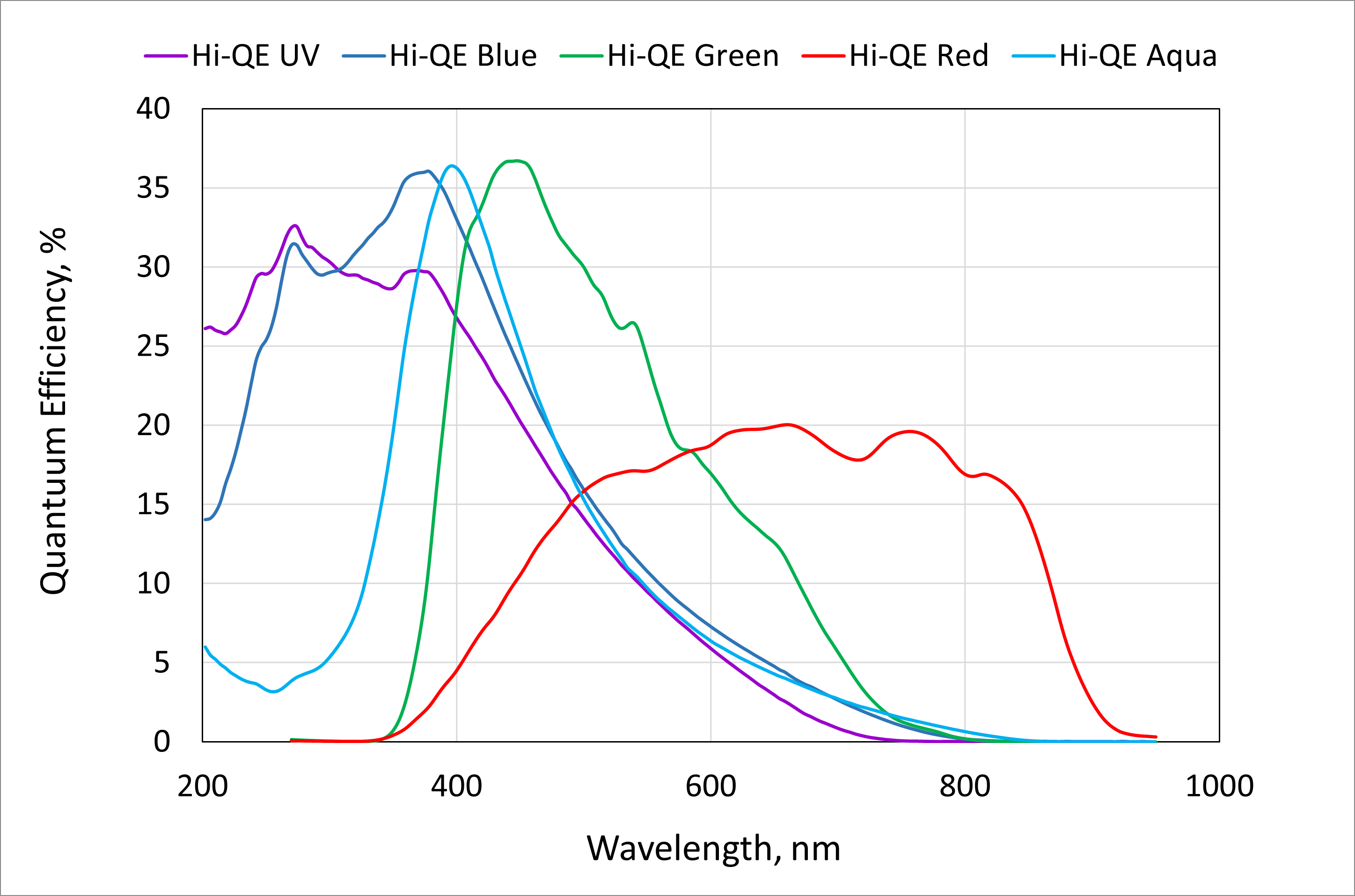}
\qquad
\includegraphics[width=.47\textwidth]{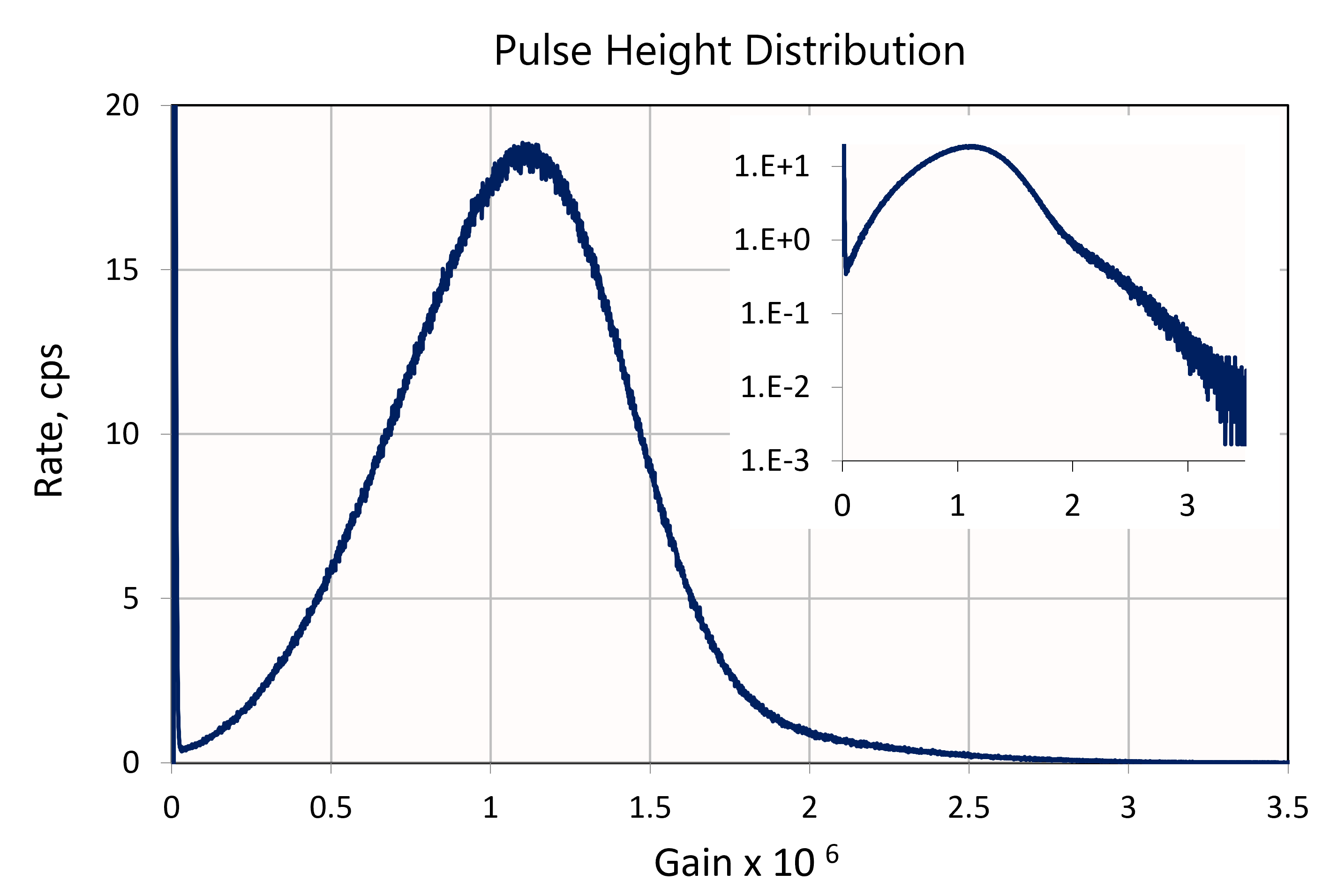}
\caption{Left: Quantum efficiency spectra for various Hi-QE series photocathodes. 
Right: Typical pulse height distribution (PHD) of the FT-8 MCP-PMT with a mean gain of about $1.1 \times 10^{6}$ (dark blue).}
\label{fig:QE}
\end{figure}

The FT-8 MCP-PMT single-photon detector is based on microchannel-plate photomultiplier vacuum tube technology with a large input window of $\varnothing$8~mm. It demonstrates excellent temporal resolution down to 10~ps~($\sigma$) and a well-defined output pulse waveform with rise/fall times below 200~ps and a pulse width of 300–-350~ps; allowing resolution of pulses separated by only a fraction of a nanosecond \cite{Ceas Lidar}. Combined with a dark count rate below 100~cps at room temperature,
low afterpulsing (about 0.1\%) \cite{Ceas Lidar},
and high quantum efficiency up to 35\% in the UV and visible spectral range
\cite{Jinst HiQE} (see also Fig.~\ref{fig:QE}, left),
the FT-8 with Hi-QE UV, Blue, Aqua, and Green photocathodes
is an ideal detector for photon-counting applications. For detection spectral range from 450~nm to 900~nm Hi-QE Red photocathodes are also available; however, they exhibit higher thermal noise, up to 50–70~kcps at room temperature.  

For high photon-rate applications, the lifetime of MCP-PMTs can be an issue. Back-stream of ions and neutral molecules desorbed from the MCP channels results in QE degradation. Typically, detector lifetime is quantified as the total charge collected at the anode over the entire life cycle. For conventional MCP, the lifetime is found to be 2--4~C/cm$^2$ \cite{Ceas Lidar}, which corresponds to 1--2 years for homogeneous illuminated input window $\varnothing$8~mm (or about 5 years for $\varnothing$18~mm) calculated with continuous operation at a pulse rate of 200~kHz and an MCP gain of $10^{6}$.  With long-lifetime MCP technology, the detector lifetime can increase by approximately a factor~10 \cite{Spie Materne}.   

Detector linearity is another factor limiting the maximum detection rate \cite{Ceas Lidar} of MCP based detectors. When the anode current reaches about 5--15\% of the strip current (scaled to the illuminated area), the depletion of the MCP charge cannot be fully compensated by recharging,  resulting in a decrease of gain and pulse amplitude. For FT-8 MCP-PMTs with the MCP gain set to $10^{6}$, saturation starts at an anode current of about 0.3~$\mu$A, corresponding to a pulse rate of approximately 2~MHz.  

Figure~\ref{fig:QE} (right) shows the PHD for typical MCP-PMT settings used in PhotonPix.
Thanks to high peak-to-valley ratio of about 50 and an FWHM/gain of 0.7--0.8 only a small fraction
of pulses have amplitude below the electronics threshold (set just above noise level) and will be lost for detection. 
Moreover, in the counting regime, when the average pulse amplitude decreases
due to saturation, the number of pulses falling below the threshold can still
remain low.
As a result, the counter reaches saturation at a later stage compared to the
linear regime \cite{Ceas Lidar}

\section{PhotonPix\texttrademark\ architecture}
\label{PhotonPix}

\begin{figure}[htbp]
\centering
\includegraphics[width=1.0\textwidth]{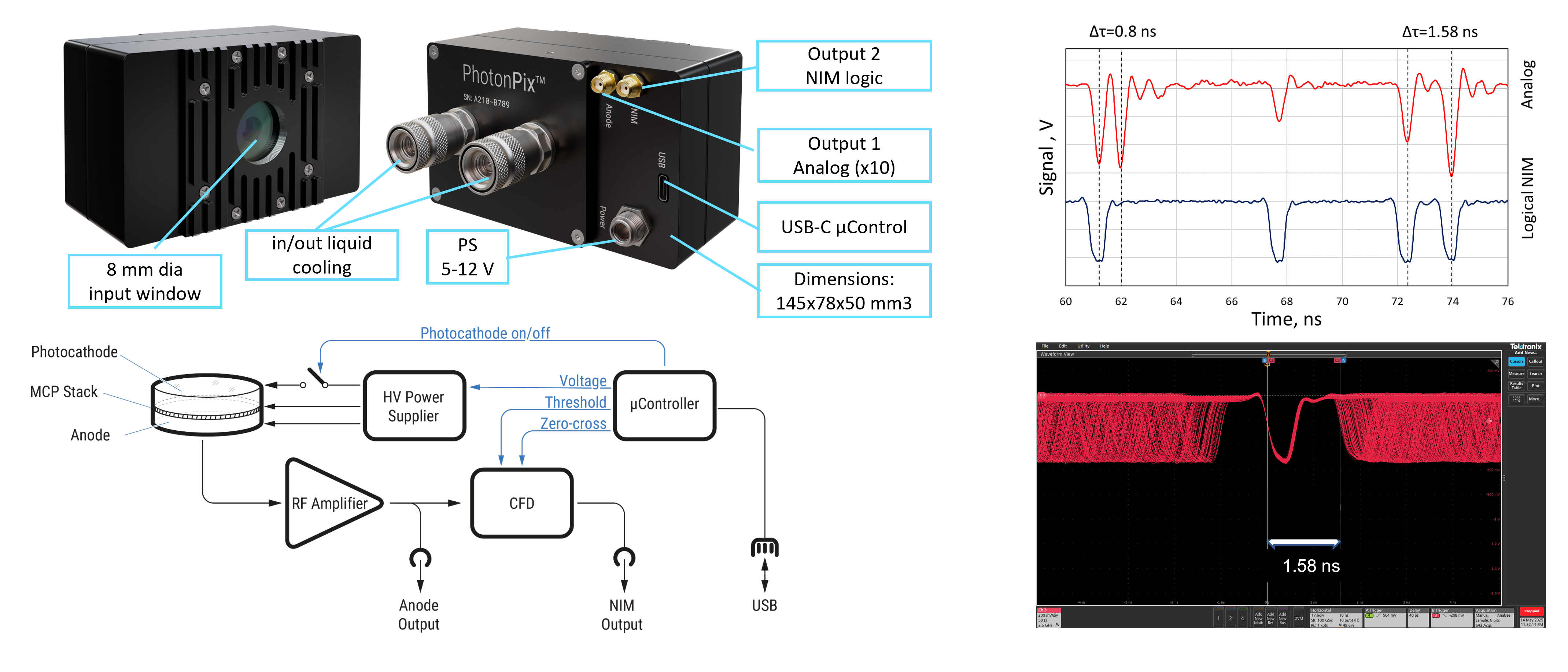}
\qquad
\caption{Left: Image of the PhotonPix (top) and schematic of the integrated photon detector module (bottom). 
Right: Waveforms from the analog and logical NIM outputs (top) and an oscilloscope trace of the NIM channel used as a trigger, integrated over several seconds, demonstrating a very low detector dead time of about 1.6~ns (bottom).}
\label{fig:PhotonPix}
\end{figure}

Implementation of electronics capable of handling high count rates with good timing accuracy is crucial to fully exploit the potential of the FT-8 MCP-PMT detector.
The PhotonPix is a plug-and-play solution, providing photon detection at the input and a logical NIM signal at the output (see Fig.~\ref{fig:PhotonPix}, left).
It is a compact detector with external dimensions of 145×78×50~mm$^3$.
PhotonPix distinguishes itself from a bare MCP-PMT by integrating all necessary electronics.
The HV divider applies voltages to three main elements: the photocathode, the MCP stack, and the anode.
The anode output is connected to a fast RF amplifier, whose signal is split by a fast active splitter into two channels (see Fig.~\ref{fig:PhotonPix}).
One signal is available to the user via a 50~$\Omega$ SMA connector labeled ``Anode'', while the second is fed into a constant fraction discriminator (CFD). The output of the discriminator is a 20~mA logical current, which produces a -0.5~V signal across a 50~$\Omega$ load.
The dead time of the PhotonPix is about 1.6~ns (see Fig.~\ref{fig:PhotonPix}, right).

The microcontroller allows computer control of the threshold and zero-crossing CFD parameters, and also of the HV power supply applied to the MCP stack to optimize parameters for measurement conditions.
The photocathode is powered independently and can be switched on or off within hundreds of milliseconds via software.
It also includes an overcurrent protection circuit to switch off the photocathode in the case of overexposure within 1~ms.

The detector assembly is cooled by an integrated Peltier element with secondary liquid cooling, reducing thermal count rates by a factor of 100 compared to room-temperature operation.
As a result, even Hi-QE Red photocathodes, which have dark rates of 50--70~kcps at room temperature, typically operate at around 200~cps in the PhotonPix detector.

\section{Ultimate timing resolution in wide photon dynamic range}
\label{Result}
\begin{figure}[htbp]
\centering
\includegraphics[width=.93\textwidth]{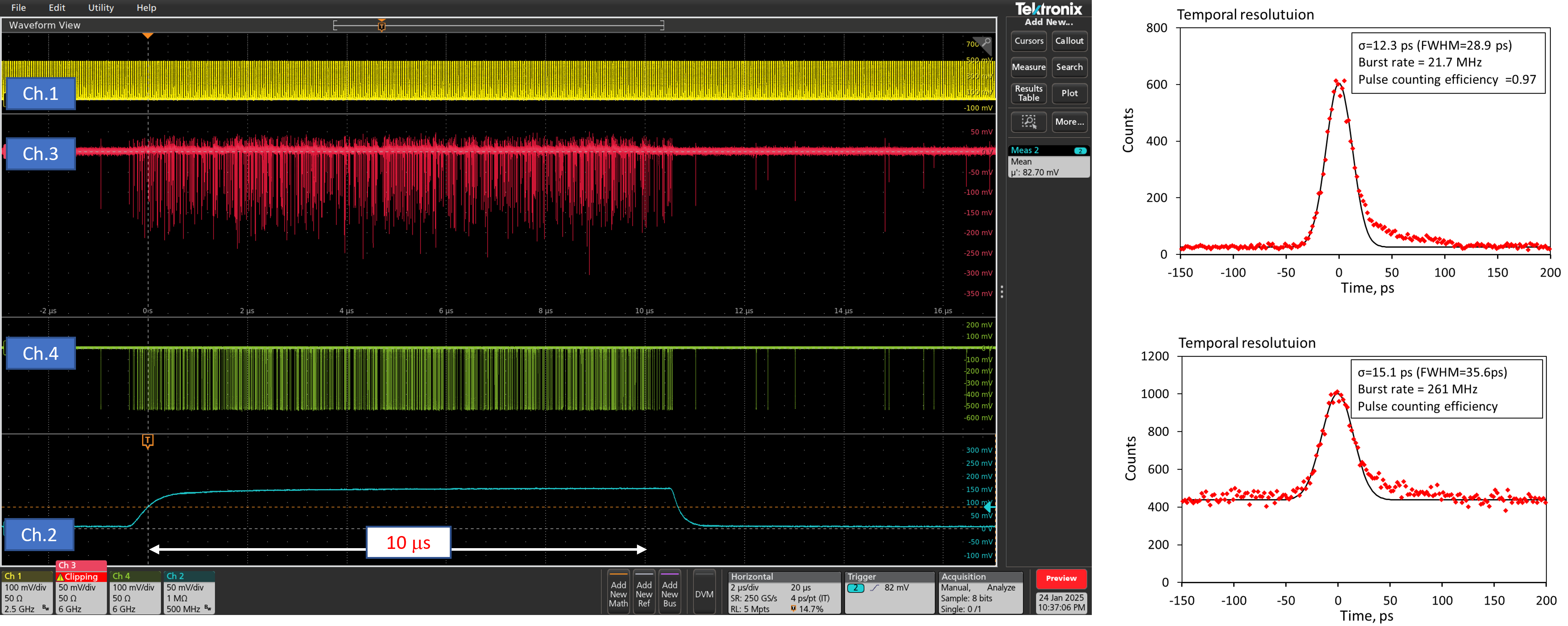}
\qquad
\caption{Left: Oscilloscope screenshot for burst-illumination measurements. Ch.~1: 64~MHz probe laser reference; Ch.~2: LED pulse reference; Ch.~3 and Ch.~4: 50~$\Omega$ analog and logical NIM outputs.
Right: Transit time spread measured with the probe laser beam at PE rates of 22~MHz (top) and 260~MHz (bottom).}
\label{fig:Osci}
\end{figure}

To characterize the detector behavior at different rates in burst mode, the PhotonPix was continuously illuminated by a probe femtosecond pulsed laser operating at a repetition rate of 64~MHz.
In addition, light from a triggered Light Emission Diode (LED) source was mixed into the beam.
The probe laser intensity was adjusted to record photoelectron (PE) pulses at about 1~MHz, ensuring single-photon detection conditions.
The LED was triggered at 10~Hz with a pulse duration of approximately 10~$\mu$s.
Under these conditions, the detector maintained linear behavior across the full range of LED intensities.
Using a 6~GHz Tektronix oscilloscope, the analog and logical outputs from the PhotonPix were recorded along with the probe laser and LED reference signals (via a fast Si photodiode) (see Fig.~\ref{fig:Osci}, left).   
The reference LED signal is used to control changes of the LED intensity and to derive PE rate at high illumination using data of LED induced PE rate recorded at low illumination settings.

To measure the transit time spread (TTS), several thousands of waveforms were acquired and analyzed by correlating pulses from the logical NIM output (within the LED-gated window) with the probe laser reference. In total, about 10--40k points were recorded for each TTS measurement. As an example, Fig.~\ref{fig:Osci} (right) shows the TTS measured at different LED intensities, corresponding to PE pulse rates of 22~MHz (top) and 260~MHz (bottom).
The TTS exhibits a slightly asymmetric shape with an extended right-side tail caused by back-scattering of photoelectrons from the MCP surface. The offset in the TTS data increases at higher LED intensities, originating from uncorrelated LED pulses.
To estimate the time jitter, the main peak was fitted with a Gaussian function.   
                    
\begin{figure}[htbp]
\centering
\includegraphics[width=.47\textwidth]{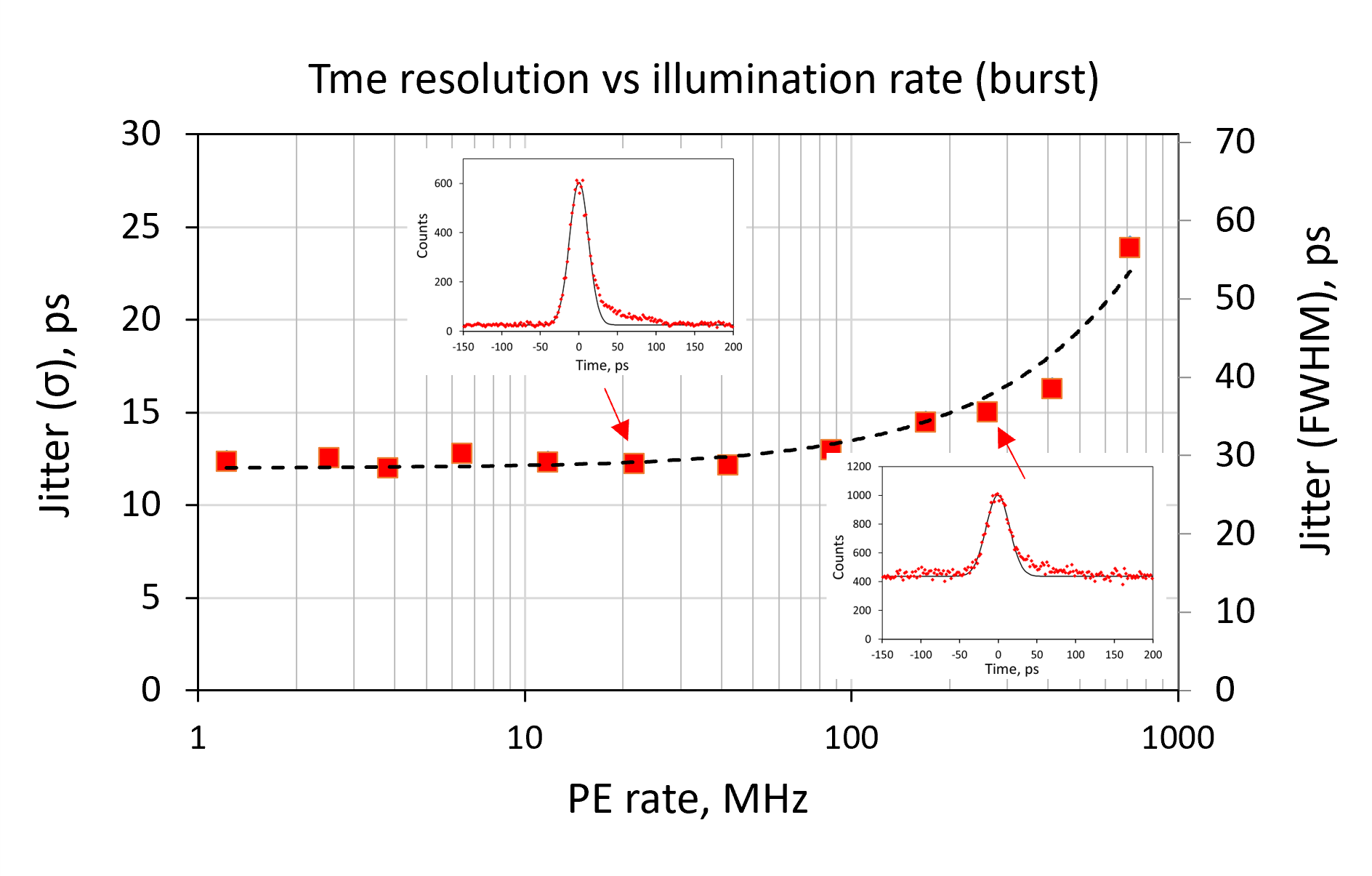}
\qquad
\includegraphics[width=.47\textwidth]{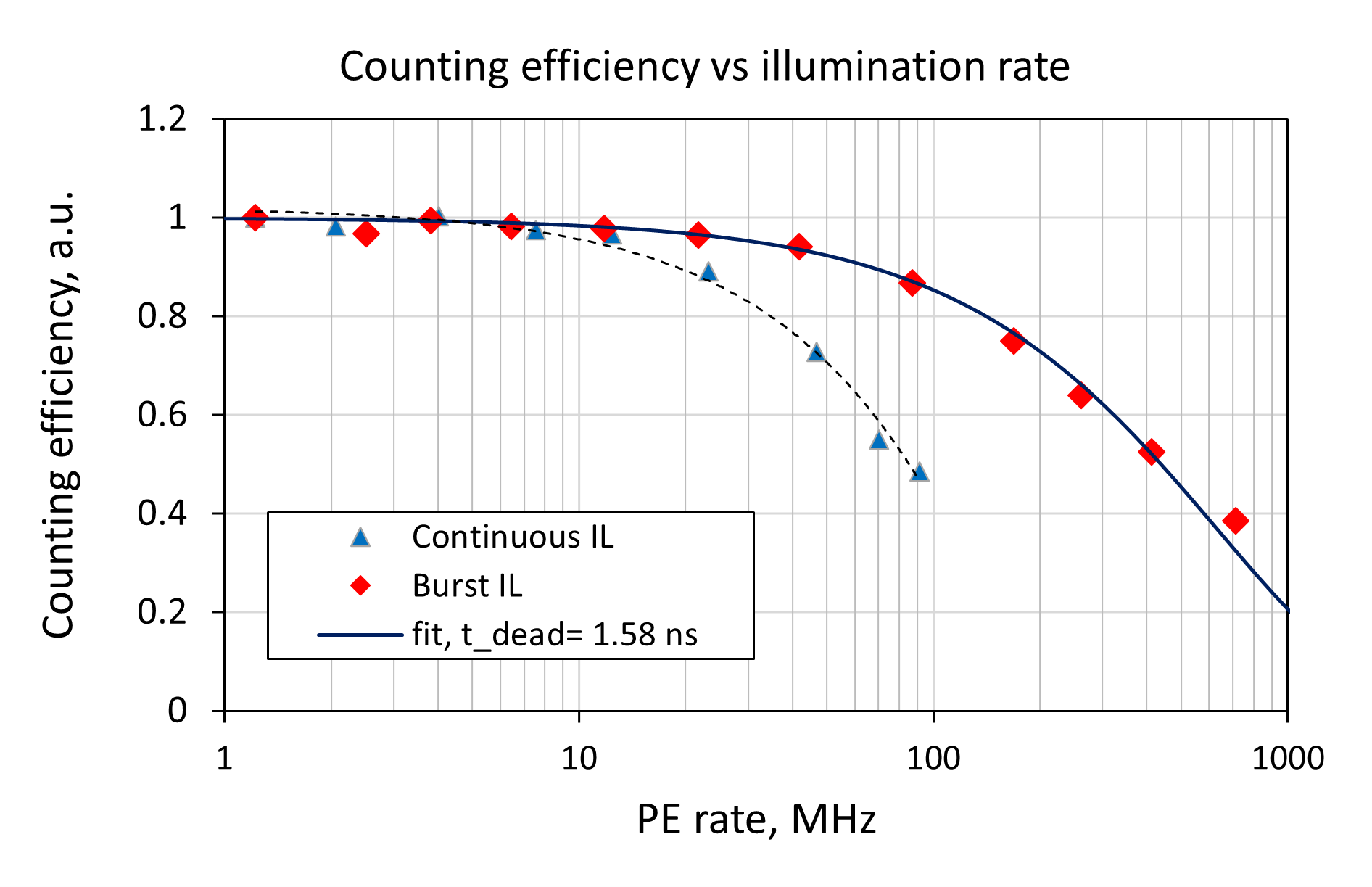}
\caption{Left: Jitter versus PE burst rate varied by adjusting the LED intensity; the dashed line shows the interpolation. 
Right: Detector counting efficiency versus PE rate measured in continuous illumination (blue, with dashed interpolation) and burst mode (red), with the black line showing the calculated curve for a detector dead time of 1.58~ns.}
\label{fig:jitter}
\end{figure}

Figure~\ref{fig:jitter} (left) shows the jitter versus PE rate recorded in burst mode illumination.
The jitter remains about 12~ps ($\sigma$) up to about 50~MHz, with a slow increase at higher rates, still staying below 25~ps at maximum PE rate of 700~MHz. The presented TTS is raw data, not corrected for oscilloscope- and analysis-induced jitter, which is estimated to be 5--6~ps ($\sigma$) in total.
For continuous LED illumination (not shown on the figure) the jitter shows almost no change across the entire measured range up to about 100~MHz for the tested sample.

\newpage
The counting efficiency (CE) is defined here as the ratio of probe-laser-induced NIM pulses within the LED window to those outside the window (which was found to be about 1 MHz for all measurements). 
Figure~\ref{fig:jitter} (right, red points) shows CE as a function of the PE burst pulse rate. 
The CE remains approximately constant up to 30~MHz and drops to 50\% at about 420~MHz.
This behavior follows the calculated model (black line in Fig.~\ref{fig:jitter}), assuming a Poisson LED light distribution and a paralyzable detector dead time of 1.58~ns.
The same figure also shows the CE for continuous LED illumination (blue points), where the CE drops to 50\% at about 90~MHz.
This reduction is caused by MCP saturation, which occurs at rates much higher than the estimated few MHz saturation limit, thanks to the well-defined PHD shape and low-noise electronics capable of setting a low counting threshold (see Sections~\ref{sec:FT8}--\ref{PhotonPix}).

In conclusion, the PhotonPix\texttrademark\ detector demonstrates ultra-high time resolution down to 10~ps.
Thanks to a short dead time of about 1.6~ns, its dynamic range extends up to approximately 1~GHz of photon flux in burst mode.
At low photon flux, detection is limited only by noise-induced dark counts, typically below 10~cps.

\end{document}